\documentclass[a4paper]{report}
\usepackage[utf8]{inputenc}
\usepackage[T1]{fontenc}
\usepackage{RJournal}
\usepackage{amsmath,amssymb,array}
\usepackage{booktabs}

\usepackage{float}

\begin{document}

\sectionhead{Contributed research article}
\volume{}
\volnumber{}
\year{}
\month{}

\begin{article}
  
\title{\pkg{diproperm}: An R Package for the DiProPerm Test}
\author{by Andrew G. Allmon, J.S. Marron, and Michael G. Hudgens}

\maketitle

\abstract{
High-dimensional low sample size (HDLSS) data sets emerge frequently in many biomedical applications. A common task for analyzing HDLSS data is to assign data to the correct class using a classifier. Classifiers which use two labels and a linear combination of features are known as binary linear classifiers. The direction-projection-permutation (DiProPerm) test was developed for testing the difference of two high-dimensional distributions induced by a binary linear classifier. This paper discusses the key components of the DiProPerm test, introduces the \pkg{diproperm} R package, and demonstrates the package on a real-world data set.
}

\section{Introduction}

Advancements in modern technology and computer software have dramatically increased the demand and feasibility to collect high-dimensional data sets. Such data possess challenges which require the creation of new and adaptation of existing statistical methods.  One such challenge is that we may observe many more predictors,  \( p \), than the number of observations,  \( n \), especially in small sample size studies.  These data structures are known as high-dimensional, low sample size (HDLSS) data sets, or $``$small  \( n \), big  \( p \) $"$ . \par

HDLSS data emerge frequently in many health-related fields. For example, in genomic studies, a single microarray experiment might produce tens of thousands of gene expressions compared to the few samples studied, often being less than a hundred \citep{Alag:2019}.\  In medical imaging studies, a single region of interest for analysis in an MRI or CT-scan image often contains thousands of features compared to the small number of samples studied \citep{Limkin:2017}.\  In pre-clinical evaluation of vaccines and other experimental therapeutic agents, the number of biomarkers measured (e.g., immune responses) may be much greater than the number of samples studies (e.g., mice, rabbits, or non-human primates) \citep{Kimball:2018}.\par

One common task in the HDLSS setting entails constructing a classifier which appropriately assigns samples to the correct class. For example, in pre-clinical studies investigators may wish to predict whether an animal survives to a certain time point based on high-dimensional biomarker data. When the data are to be partitioned into two classes, binary linear classifiers have been shown to be especially useful in HDLSS settings and preferable to more complicated classifiers because of their ease of interpretability \citep{Aoshima:2018}. However, linear classifiers may find spurious linear combinations in HDLSS settings \citep{Marron:2007}. That is, a binary linear classifier may find, for two identical high-dimensional distributions, a linear combination of features which incorrectly suggests the two classes are different. Thus, it is important to assess whether a binary linear classifier is detecting a statistically significant difference between two high-dimensional distributions. \par


\section{DiProPerm}

The direction-projection-permutation (DiProPerm) test\ was developed to test whether or not a binary linear classifier detected a difference between two high-dimensional distributions \citep{Wei:2016}. DiProPerm uses one-dimensional projections of the data based on the binary linear classifier to construct a univariate test statistic, and then permutes class labels to determine the sampling distribution of the test statistic under the null.  Importantly, the DiProPerm test is exact, i.e., the type I error is guaranteed to be controlled at the nominal level for any sample size.  \par

To better understand the mechanics of DiProPerm, let  \( X_{1}, \ldots ,X_{n} \sim F_{1} \)  and  \( Y_{1}, \ldots ,Y_{m} \sim F_{2} \)  be independent random samples of  \( p \)  dimensional random vectors from multivariate distributions  \( F_{1} \)  and  \( F_{2} \)  where  \( p \) may be larger than \(m\) and \(n\).\  The DiProPerm tests \par

 \[ H_{0}:F_{1}=F_{2} \text{ \  versus \  } H_{1}:F_{1} \neq F_{2} \] 

\noindent The general idea of the DiProPerm test can be explained in three steps: \par

\begin{enumerate}
	\item Direction:  find the normal vector to the separating hyperplane between two samples after training a binary linear classifier\par

	\item Projection:  project data on to the normal vector and calculate a univariate two-sample statistic\par

	\item Permutation:  conduct a permutation test using the univariate statistic as follows:\par

\begin{enumerate}
	\item Permute class membership after pooling samples \par

	\item Re-train binary classifier and find the normal vector to the separating hyperplane \par

	\item Recalculate the univariate two sample statistic \par
	
	\item Repeat a-c multiple times (say 1000) to determine the sampling distribution of the test statistic under the null \( H_{0}\) \par
	
	\item Compute p-value by comparing the observed statistic to the sampling distribution \par
\end{enumerate}
\end{enumerate}\par

Different binary linear classifiers may be used in the first step of DiProPerm. Linear discriminant analysis (LDA), particularly after conducting principal component analysis (PCA), is one possible classifier for the direction step.  However, using LDA with PCA in the HDLSS setting has some disadvantages, including lack of interpretability, sensitivity to outliers, and tendency to find spurious linear combinations due to a phenomenon known as data piling \citep{Aoshima:2018,Marron:2007}.  Data piling occurs if data are projected onto some projection direction and many of the projections are the same, or piled on one another.  The support vector machine (SVM)  is a another popular classifier \citep{Hastie:2001}. The SVM finds the hyperplane that maximizes the minimum distance between data points and the separating hyperplane. However, the SVM can also suffer from data piling in the HDLSS setting.  To overcome data piling, the distance weighted discrimination (DWD) classifier was developed \citep{Marron:2007}.\ \ The DWD classifier finds the separating hyperplane minimizing the average inverse distance between data points and the hyperplane.  The DWD performs well in HDLSS settings with good separation and is more robust to data piling. \par

In the second step of DiProPerm, a univariate statistic is calculated using the projected values on to the normal vector to the separating hyperplane from the first\ step.  Suppose \( x_{1}, \ldots ,x_{n} \) and \( y_{1}, \ldots ,y_{m} \) are the projected values from samples  \( X_{1}, \ldots ,X_{n} \) and \( Y_{1}, \ldots ,Y_{m} \),\ respectively.  One common choice for the univariate test statistic for DiProPerm includes the difference of means statistic, \( |\bar{x}-\bar{y}| \). Other two-sample univariate statistics such as the two-sample t-statistic or difference in medians are also possible for use with the DiProPerm.\ \par

The last step of DiProPerm\ entails determining the distribution of the test statistic under the null.  In this step, the two samples are pooled, class labels are permuted, then a univariate statistic is calculated. Repeat this process multiple times (say 1000) to determine the sampling distribution of the test statistic under the null \( H_{0} \).  P-values are then calculated by the proportion of statistics higher than the original value. \par

When the DiProPerm test is implemented using the DWD classifier, it is common practice to look at the loadings of the DWD classifier \citep{An:2016,Nelson:2019}.\ \ The\ DWD loadings represent the relative contribution of each variable to the class difference.  A higher absolute value of a variable’s loading indicates a greater contribution for that variable to the class difference.  Combining the use of the DiProPerm and evaluation of the DWD loadings in applications can provide insights into high-dimensional data and be used to generate rational hypotheses for future research.  \par

The DiProPerm test has been used in several areas of biomedical research including osteoarthritis and neuroscience \citep{An:2016,Bendich:2016,Nelson:2019}. However, currently there does not exist an R package which implements DiProPerm. Therefore we developed \pkg{diproperm}, a simple, free, publicly available R software package to analyze data from two high-dimensional distributions. \pkg{diproperm} displays diagnostic plots for a specified univariate statistic and calculates p-values for the DiProPerm test.\  The loadings for the binary linear\ classifier are also available for display in order from highest to lowest relative to their contribution toward the separation of the two distributions.   \par


\section{The diproperm package}

The \pkg{diproperm} package is comprised of three functions: \par

\begin{itemize}
	\item \code{DiProPerm()}: Conducts DiProPerm test\par

	\item \code{plotdpp()}: Plots diagnostics from the DiProPerm test\par

	\item \code{loadings()}: Returns the variable indices with the highest loadings in the binary classification. The absolute values of the loading values indicate a variable's relative contribution toward the separation between the two classes.
\end{itemize} \par

\subsection{diproperm example}

The example below creates a Gaussian data set containing 100 samples, 2 features, clustered around 2 centers with a standard deviation of 2.  The class labels are then re-classified to -1 and 1 to match the input requirements of \code{DiProPerm()}. The DiProPerm test is then conducted using the DWD classifier, the mean difference univariate statistic, and 1000 permutations. The results from \code{DiProPerm()} are then displayed with \code{plotdpp()}.  Last, the top five indices of the highest absolute loadings are listed.

\begin{example}
  devtools::install_github("elbamos/clusteringdatasets")
  library(clusteringdatasets)
    
  cluster.data <- make_blobs(n_samples = 100, n_features = 2, centers = 2, cluster_std = 2)
    
  X <- cluster.data[[1]]
  y <- cluster.data[[2]]
  y[y==2] <- -1
    
  dpp <- DiProPerm(X,y,B=1000,classifier = "dwd",univ.stat = "md")

  plotdpp(dpp)
    
  loadings(dpp,loadnum = 5)
\end{example}

\subsection{Description}

The main function to be called first by the user is \code{DiProPerm()}, which takes in an \(n \times p \)  data matrix and a vector of  \( n \)  binary class labels both\ provided by the user.  Factor variables for the data matrix must be coded as 0/1 dummy variables and the class labels for the vector of binary class labels must be coded as -1 and 1.\  By default the \code{DiProPerm()} uses the DWD classifier, the mean difference as the univariate statistics, and 1000 balanced permutations.  The permutations are balanced in the sense that after relabeling, the new -1 group contains \(n/2\) members from the original -1 group and \(n/2\) members not from the original -1 group.  \code{DiProPerm()} implements DWD from the \code{genDWD} function in the \CRANpkg{DWDLargeR} package \citep{DWDLargeRPackage,Lam:2018}. The penalty parameter, \code{C}, in the \code{genDWD} function is calculated using the \code{penaltyParameter} function in \CRANpkg{DWDLargeR}.  More details on the algorithm used to compute \code{genDWD} and \code{penaltyParameter} can be found in \citet{Lam:2018}. Another option included in \code{DiProPerm()} for the binary linear classifier is "md", mean difference direction.\  Users can also implement an unbalanced, randomized permutation design if desired.\  \code{DiProPerm()} uses parallel processing to delegate computation to the number of cores on the user’s computer for increased efficiency.\  \code{DiProPerm()} returns a list of the observed data matrix, vector of observed class labels, observed test statistic, projection scores used to compute the observed test statistic, the loadings of the binary classification, the z-score, cutoff value for an  \(  \alpha  \)  level of significance, the p-value for the DiProPerm test, a list of each permutation’s projection scores and permuted class labels, and a vector of permuted test statistics the size of the number of permutations used. \par

After fitting the \code{DiProPerm()}, the user can use \code{plotdpp()} to create a panel plot for assessing the diagnostics of the DiProPerm\ test.  \code{plotdpp()} takes in a DiProPerm\ list and the user may specify which diagnostics they would like to display.  By default, \code{plotdpp()} displays a facet plot with the observed score distribution, the projection score distribution of the permutation with the smallest test statistic value, the projection score distribution of the permutation with the largest test statistic value, and the test statistic permutation distribution. For the permutation distribution plot, the z-score, cutoff value, observed test statistic and p-value\ are displayed on the graph.  Larger, individual graphs may be displayed by using the plots option in \code{plotdpp()}.\  Additional graphs include the projection score distributions for the first and second permutations.\  The diagnostic plots show the user the characteristics of their data and facilitate the visual assessment of the separation of the two high-dimensional distributions being tested. \par

Lastly, after calling the \code{DiProPerm()}, the user may call the \code{loadings()} function.  The \code{loadings()} function returns the variable indices in the data matrix which have the highest absolute loadings in the binary classification. Higher absolute loading values indicate a greater contribution for a particular variable toward the separation between the two classes.\  By default, \code{loadings()} returns the indices for all variables sorted by their absolute loading value. Therefore, the top variable index is the variable which contributes the most toward the separation of the two classes and the last variable is the one which contributes the least. The user may also change the number of loadings displayed. \par


\section{Application}

To illustrate use of the \pkg{diproperm} package, consider the mushrooms data set which is freely available from the UCI Machine Learning Repository \citep{Dua:2019} and within \pkg{diproperm}. This data set includes various characterizations of  \( 23\) species of gilled mushrooms in the Agaricus and Lepiota family.  Each mushroom species is labeled as either definitely edible or poisonous/unknown.  There are \(n=8124\) mushrooms total, and \(p=112\) binary covariates coded as 0/1 corresponding to 22 categorical attributes. Below we demonstrate the \pkg{diproperm} package functionality using data from the first \(n=50\) mushrooms in the data set.\par

\subsection{Step\ 1:  Load and clean the data}\par

\begin{example}
  devtools::install_github("allmondrew/diproperm")
  library(diproperm)
  data(mushrooms)
\end{example}
The above code installs the \pkg{diproperm} package and loads the mushroom data into R.\  Now let’s check the structure of the data to make sure it is compatible with \code{DiProPerm()}.\par

\begin{example}
  dim(mushrooms$X)
      [1] 112 8124
  
  table(mushrooms$y)
      -1 1
      4208 3916
\end{example}
The vector of class labels must be -1 or 1 for \code{DiProPerm()} which is the case for this data; however, the data set is in  \( p \times n \) \ format.  For \code{DiProPerm()}, the dataset must be in  \( n \times p \) \ format.\  This can be done using the transpose function from the \CRANpkg{Matrix} package in R \citep{MatrixPackage}.  After taking the transpose, we subset the data and vector of class labels to the first 50 observations and store the results. \par


\begin{example}
  X <- Matrix::t(mushrooms$X)
  X <- X[1:50,]
  y <- mushrooms$y[1:50]
\end{example}

\subsection{Step 2:\  Conduct DiProPerm}\par

Now that the data is in the proper format the call to \code{DiProPerm()} is as follows\par

\begin{example}
  dpp <- DiProPerm(X=X,y=y,B=1000)
  
     Algorithm stopped with error 2.35e-08
     sample size =  50, feature dimension = 112
     positive sample =  12, negative sample =  38
     number of iterations =  51
     time taken = 0.39
     error of classification (training) = 0.00 (
     primfeas = 3.49e-10
     dualfeas = 0.00e+00
     relative gap = 2.89e-07
\end{example}
Characteristics of the DWD algorithm used to find the solution for the observed data are displayed by \code{DiProPerm()}.\  The algorithm took 51 iterations and 0.39 seconds to converge to the tolerance threshold with a zero percent classification error on the training data set. The runtime for 1000 permutations was less than 3 minutes on a four-core machine but would be faster on a machine with more cores. The \code{dpp} object stores the output list from \code{DiProPerm()}\ described in the package.  Storing the information allows us to plot the diagnostics in the next step.\  \par

\subsection{Step 3:\  Plot diagnostics}\par

\begin{example}
  plotdpp(dpp)
\end{example}

\begin{figure}[H]
  \centering
  \includegraphics[width=\textwidth,keepaspectratio]{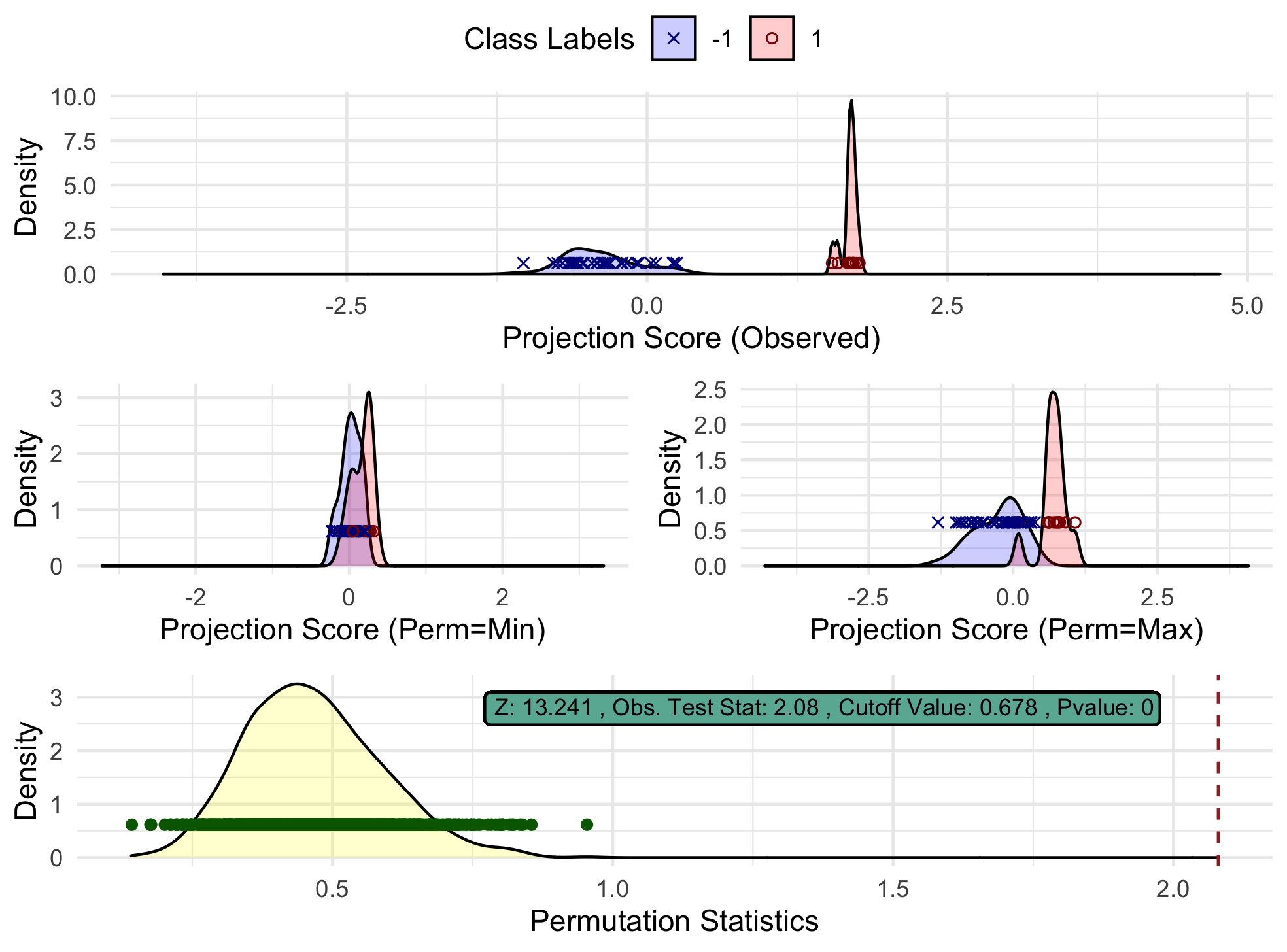}
  \caption{The diagnostic plot from \code{plotdpp()} for the mushrooms data set.\ The top graph is the observed projection score distribution of the two classes, the two middle graphs are the projection score distributions of the permutation with the smallest and largest test statistic value, and the bottom graph is the test statistic permutation distribution with the observed statistic value marked by the red dotted line.}
  \label{figure:plotdpp}
\end{figure}

\noindent Figure \ref{figure:plotdpp} displays the default diagnostics for a DiProPerm list.\  From the observed projection score distribution, one can see clear separation between the two classes.\ \ Also, from the projected score distributions of the permutations which yield the smallest and largest test statistic, we see the score distributions overlap well so there is some visual justification that the distributions in the observed plot are truly different. Lastly, the bottom plot shows the sampling distribution under the null is located around 0.4 while the observed test statistic is greater than 2.  Each individual plot can also be output by the following set of commands\par

\begin{example}
  plotdpp(dpp,plots="obs")
  plotdpp(dpp,plots="min")
  plotdpp(dpp,plots="max")
  plotdpp(dpp,plots="permdist")
\end{example}

The permutation p-value in Figure \ref{figure:plotdpp} suggests that the two high-dimensional distributions of mushroom attributes are indeed different between the two classes. Also displayed is a z-score, calculated by fitting a Gaussian distribution to the test statistic permutation distribution. The mushroom data z-score 13.2 indicates the observed test statistic is approximately 13 standard deviations from the expected value of the test statistic under the null. Finally, the cutoff value 0.678 is displayed, corresponding to the critical value for a hypothesis test at the 0.05 significance level. \par

\subsection{Step 4:\  Examine loadings}\par

In order to assess which variables contributed most toward the separation in step 3 we can print the top five contributors with the code\par


\begin{example}
  loadings(dpp,loadnum = 5)
  
      index sorted_loadings
      29    0.5395016
      37    0.3170037
      36    -0.2481763
      111   0.2228389
      20    -0.2087244
\end{example}

The\ top five contributors toward the separation seen in the observed distribution in Figure \ref{figure:plotdpp} are indices 29, 37, 36, 111, and 20. These indices correspond to a pungent odor, narrow gill size, broad gill size, urban habitat, and yellow cap color, respectively. These results are similar to previous analyses which have also found odor, gill size, habitat, and cap color predictive of mushroom edibility \citep{Pinky2019,Wibowo2018}. \par


\section{Summary}
Binary linear classifiers can suffer from finding spurious separating directions in the HDLSS setting, i.e., data may be sampled from two identical distributions but the binary linear classifier may find a linear combination of features such that the two classes appear to be very different.  The DiProPerm test was created to test whether or not the separation induced by the binary linear classifier is truly separate or just a result of over-fitting.  The \pkg{diproperm} package allows the user to visually assess and empirically test if there is a difference between the high-dimensional distributions of the two classes and, if so, evaluate the key features contributing to the separation between the classes. \par

\section{Acknowledgments}
This work was supported by the National Institute of Allergy and Infectious Diseases (NIAID) of the National Institutes of Health (NIH) under award number R37AI054165. The content is solely the responsibility of the authors and does not necessarily represent the official views of the NIH. 
\par

\bibliography{allmon}

\address{Andrew G. Allmon\\
  University of North Carolina at Chapel Hill \\
  Department of Biostatistics \\
  \email{dallmon@email.unc.edu}}

\address{James S. Marron\\
  University of North Carolina at Chapel Hill \\
  Department of Biostatistics \\
  \email{marron@unc.edu}}

\address{Michael G. Hudgens\\
  University of North Carolina at Chapel Hill \\
  Department of Biostatistics \\
  \email{mhudgens@email.unc.edu}}

\end{article}

\end{document}